\documentclass[11pt]{article}

\PassOptionsToPackage{table,dvipsnames}{xcolor}

\usepackage[preprint]{acl}

\usepackage{times}
\usepackage{latexsym}
\usepackage[T1]{fontenc}
\usepackage[utf8]{inputenc}
\usepackage{microtype}
\usepackage{inconsolata}

\usepackage{amsmath}
\usepackage{amssymb}

\usepackage{float}
\usepackage{algorithm}
\usepackage{algpseudocode}

\usepackage{graphicx}
\graphicspath{{figures/}}

\usepackage{booktabs}
\usepackage{multirow}
\usepackage{makecell}
\usepackage{colortbl}

\definecolor{whisfusionblue}{HTML}{DCE6F2}
\definecolor{baselinebrown}{HTML}{F2E6D9}
\definecolor{diffyellow}{HTML}{FFF2CC}

\title{Whisfusion: Parallel ASR Decoding with Masked Diffusion}

\author{\mdseries
  Taeyoun Kwon$^{1}$, Junhyuk Ahn$^{1}$, Taegeun Yun$^{2}$, Heeju Jwa$^{1}$, Yoonchae Choi$^{1}$, \\
  Siwon Park$^{2}$, Jongchan Kim$^{1}$, Hyungon Ryu$^{3}$, Hyuk-Jae Lee$^{1}$, Nam-Joon Kim$^{1}$
  \\[0.3em]
  $^{1}$Seoul National University \quad
  $^{2}$Soongsil University \quad
  $^{3}$NVIDIA Corporation \\[0.2em]
  \texttt{ty8352@snu.ac.kr}, \texttt{knj01@snu.ac.kr}
}

\hypersetup{
  pdftitle={Whisfusion: Parallel ASR Decoding with Masked Diffusion},
  pdfauthor={Taeyoun Kwon, Junhyuk Ahn, Taegeun Yun, Heeju Jwa, Yoonchae Choi, Siwon Park, Jongchan Kim, Hyungon Ryu, Hyuk-Jae Lee, Nam-Joon Kim},
  pdfkeywords={automatic speech recognition, masked diffusion, non-autoregressive, parallel decoding}
}

\begin{document}
\maketitle

\begin{abstract}
Autoregressive (AR) encoder-decoder models dominate high-quality
multilingual ASR, but their left-to-right decoders make inference
latency scale with transcript length. A natural alternative, CTC-style non-autoregressive
(NAR) systems avoid this bottleneck but their conditional
independence assumption sacrifices transcript-level generative
modeling. Masked diffusion language models (e.g., LLaDA, MDLM)
offer a competitive NAR text-generation approach.
We ask whether such models can bring NAR ASR into the accuracy
regime of strong AR ASR systems while removing the left-to-right
bottleneck. We propose
Whisfusion, which trains a dedicated masked diffusion decoder from
scratch on top of frozen Whisper-large-v3 audio embeddings, denoising
masked transcripts in just a few steps. We train on $\sim$68k hours of
11-language speech with high-mask specialization to align training
with the fully masked starting point of inference, and decode via
Parallel Diffusion Decoding. Whisfusion surpasses Whisper-large-v3 on
group-average accuracy across English, European, and CJK benchmarks,
while running 4--5$\times$ faster, additionally surpassing
Whisper-turbo in both accuracy and throughput. It reaches
accuracy competitive with Canary and Qwen3-ASR while running
3--7$\times$ faster. These results establish masked diffusion as a
Pareto-competitive non-autoregressive paradigm for high-throughput
multilingual transcription.
Code and model weights are available at
\url{https://github.com/taeyoun811/Whisfusion}.
\end{abstract}

\section{Introduction}
\label{sec:intro}

Autoregressive encoder-decoder and audio-language models dominate
high-quality multilingual automatic speech recognition (ASR). Systems such as Whisper, Canary, and Qwen3-ASR
\citep{radford2023whisper,canary2025,qwen3asr2026} achieve strong
transcription accuracy, multilingual coverage, and robustness across
diverse domains by combining large-scale speech pretraining with
powerful text generation decoders.

However, their left-to-right decoding mechanism creates a structural
latency bottleneck. While the encoder produces audio
embeddings for the full utterance, the decoder still serializes
transcript generation token by token, repeatedly attending to the same
encoded audio while producing $y_1, y_2, \ldots, y_T$. This makes
decoding latency grow with transcript length and causes the decoder to
dominate inference time as transcripts become longer or throughput
requirements increase.

A natural alternative is non-autoregressive ASR, most notably CTC-based
recognition \citep{graves2006ctc}, which parallelizes frame-level
alignment to avoid token-by-token generation. However, this efficiency
comes from a different modeling assumption: CTC primarily parallelizes
acoustic-to-token alignment rather than transcript-level generation.
Its conditional independence structure can limit long-range linguistic
modeling, and refinement-based NAR methods such as Mask-CTC
\citep{higuchi2020maskctc} improve over one-shot CTC decoding but still
remain anchored to alignment-derived hypotheses. Existing NAR ASR
thus occupies a fast but weaker accuracy regime than modern AR models.

In text generation, masked diffusion and diffusion language models
\citep{austin2021d3pm,sahoo2024mdlm,nie2025llada} provide an
alternative to left-to-right generation. Rather than
generating one token at a time, they start from a masked sequence
and iteratively denoise positions in parallel. We argue that ASR is a particularly natural setting for this
approach: the acoustic condition is already available before generation
begins, so a decoder can refine a masked transcript while attending to
the full audio embeddings.

We introduce Whisfusion, a speech-conditioned masked diffusion
framework for parallel ASR decoding. Whisfusion combines a frozen
Whisper-large-v3 encoder with a from-scratch masked diffusion decoder
whose every block attends to audio embeddings through cross-attention.
We train it with high-mask specialization to align training with fully
masked inference, and decode with Parallel Diffusion Decoding (PDD),
minimum Bayes risk (MBR) consensus selection, and optional Whisper
likelihood reranking.
Across English, European, and CJK benchmarks, Whisfusion shows that
non-autoregressive ASR can enter the accuracy regime of modern
autoregressive ASR while retaining the high-throughput advantage of
fixed-step parallel inference.

Our contributions are threefold: (i) we introduce Whisfusion, a
speech-conditioned masked diffusion ASR model that replaces
left-to-right token generation with fixed-step transcript denoising,
grounding a from-scratch non-autoregressive decoder in frozen
Whisper-large-v3 audio embeddings through cross-attention;
(ii) we develop a training and decoding recipe that aligns
masked-diffusion training with the inference trajectory, combining
high-mask specialization, random-remask Parallel Diffusion Decoding,
MBR consensus selection, and optional Whisper likelihood reranking;
and (iii) we evaluate Whisfusion across English, European, and CJK
benchmark groups, showing that it surpasses Whisper-large-v3 on group
averages and approaches the strongest AR systems such as Qwen3-ASR and
Canary, while retaining substantially higher throughput.

\section{Background and Related Work}
\label{sec:related}

\subsection{Non-Autoregressive ASR}
\label{sec:related-ar}

While autoregressive encoder-decoder systems remain the accuracy
frontier for multilingual ASR
\citep{radford2023whisper,canary2025,qwen3asr2026}, a parallel line
of work has explored non-autoregressive (NAR) routes around
left-to-right decoding. CTC \citep{graves2006ctc} parallelizes
frame-level acoustic-to-token alignment under per-frame conditional
independence, making it the dominant fast-decoding approach but
limiting transcript-level joint modeling. Refinement-based variants
relax this assumption: Mask-CTC \citep{higuchi2020maskctc}
iteratively rewrites low-confidence CTC outputs via masked token
prediction, and Imputer \citep{chan2020imputer} generates transcripts
through learned insertion operations. Closely related NAR text
generation from machine translation (CMLM
\citep{ghazvininejad2019cmlm} and the Levenshtein Transformer
\citep{gu2019levt}) established masked refinement and edit-based
decoding. These
methods rely on alignment-derived hypotheses, edit operations, or
fixed-iteration mask-fill heuristics. Masked diffusion offers a
different formulation---principled fixed-step transcript-level
denoising under a stochastic noise process---which we review next.

\subsection{Masked Diffusion Language Models}
\label{sec:related-mdlm}

Discrete diffusion language models, especially masked variants,
have been proposed as alternatives to autoregressive text generation.
D3PM \citep{austin2021d3pm} introduced diffusion processes over
discrete token spaces, and SEDD \citep{lou2024sedd} and MDLM
\citep{sahoo2024mdlm} sharpened the training objective for discrete
and masked diffusion, respectively. SMDM and LLaDA demonstrate that
masked diffusion scales to large language models: LLaDA trains an 8B
model from scratch and SMDM characterizes masked-diffusion scaling
laws \citep{nie2025smdm,nie2025llada}. Extending this approach to audio-conditional generation requires
grounding each denoising step in speech embeddings rather than
text-only context, a problem addressed by the diffusion ASR systems
reviewed next.

\subsection{Diffusion and dLLM-based ASR}
\label{sec:related-diff-asr}

Diffusion-style ASR systems vary both the noise process and the
decoder: TransFusion \citep{baas2022transfusion} and FDDM
\citep{yeh2024fddm} use Gaussian and discrete noise processes,
respectively, and Drax \citep{navon2025drax} uses discrete flow
matching with a frozen Whisper-large-v3 encoder, serving as a strong
public diffusion/flow baseline. More recently, MDM-ASR
\citep{yen2026mdmasr} replaces left-to-right decoding with
audio-conditioned masked diffusion and introduces training and
sampling strategies to reduce the train--inference mismatch. A
separate line of work connects speech encoders to large pretrained
diffusion language models: Whisper-LLaDA \citep{whisperllada2026}
couples a Whisper encoder with LLaDA, and dLLM-ASR
\citep{dllmasr2026} performs a similar adaptation. Whisfusion differs
by training a from-scratch, ASR-specific masked diffusion decoder at
multilingual scale, evaluated against AR, CTC, diffusion/flow, and
dLLM baselines under a unified protocol.

\begin{figure*}[!t]
  \centering
  \includegraphics[width=0.95\textwidth]{Fig_1.pdf}
  \caption{\textbf{Whisfusion architecture and masked-diffusion
  decoding.} (a) A trainable masked-diffusion decoder is conditioned
  on the frozen Whisper encoder's audio embeddings via cross-attention
  in every block. (b) Starting from a masked sequence, Whisfusion
  performs fixed-step denoising: at each step, the decoder predicts
  all currently masked transcript positions in parallel while
  attending to audio embeddings.}
  \label{fig:arch}
\end{figure*}

\section{Whisfusion}
\label{sec:method}

Whisfusion reformulates ASR decoding as speech-conditioned masked
transcript denoising. A frozen Whisper-large-v3 encoder produces
audio embeddings, and a trainable masked diffusion decoder iteratively
refines a masked transcript while attending to those embeddings
(Figure~\ref{fig:arch}).

\subsection{Speech-Conditioned Masked Diffusion Decoder}
\label{sec:method-arch}

The decoder is a from-scratch masked diffusion transformer with 24
layers, hidden size 1280, and 20 attention heads (approximately
$830$M trainable parameters). We match the decoder
hidden size to the Whisper-large-v3 encoder output dimension, which
simplifies the cross-attention interface between speech and text
representations. Following recent masked diffusion language models
\citep{sahoo2024mdlm,nie2025llada}, the decoder uses a modern
LLaMA-style stack \citep{touvron2023llama} with RMSNorm
\citep{zhang2019rmsnorm}, SwiGLU \citep{shazeer2020glu} feed-forward
layers, RoPE \citep{su2021roformer}, QK-normalization
\citep{henry2020qknorm}, bias-free projections, and untied
input/output embeddings. Each decoder block consists of
bidirectional self-attention over the partially masked transcript,
cross-attention to the audio embeddings, and a feed-forward layer.
Unlike an autoregressive decoder, the self-attention is non-causal,
allowing the model to condition on both left and right textual context
during denoising. We use the Whisper multilingual tokenizer and add a
single mask token for diffusion training and inference. Full architectural
hyperparameters are listed in Appendix~\ref{app:arch}.

\subsection{Training with High-Mask Specialization}
\label{sec:method-train}

We train Whisfusion with the standard masked diffusion objective.
Given an audio input $x$, its transcript $y_0$, and audio embeddings
$c$ from the frozen Whisper encoder, we sample a mask ratio $t$ and
corrupt the transcript into $y_t$ by replacing a subset of transcript
tokens with the mask token. The decoder is trained to recover the
original transcript tokens at masked positions, conditioned on both the
partially observed transcript $y_t$ and the audio embeddings $c$.
Prompt tokens such as language and transcription indicators are
preserved during masking, while padding positions are excluded from
the loss.

ASR differs from open-ended text generation in how errors are judged.
In text generation, an early denoising error can often be absorbed
into a fluent and coherent continuation. In ASR, however, denoising
is constrained by a single audio-conditioned transcript: a token that
is plausible in text context can still be wrong if it is not supported
by the speech signal. This makes the high-mask regime especially
important. When inference starts from a fully masked transcript,
errors made in the first denoising step can become part of the
unmasked context in later steps and bias subsequent refinements.

We therefore use a two-stage mask-range narrowing recipe. In Stage 1,
we train with the full masking range, $t \sim U(0, 1)$, following
standard masked diffusion training and teaching the model general
audio-conditioned infilling. In Stage 2, we resume from the Stage 1
model but sample only high mask ratios, $t \sim U(0.7, 1.0)$. This
keeps the same objective while specializing the model for the
inference-critical regime where little or no textual context is
available. By improving high-mask initialization, Stage 2 enables
Whisfusion to use a small fixed number of denoising steps at inference.
Per-stage optimizer, learning rate, batch size, and token budget are
reported in Appendix~\ref{app:train}. The two-stage training procedure
is given as Algorithm~\ref{alg:training} in Appendix~\ref{app:algos},
and the precise loss form is in Appendix~\ref{app:objective}.

\subsection{Parallel Diffusion Decoding and Candidate Selection}
\label{sec:method-pdd}

\begin{figure}[t]
  \centering
  \includegraphics[width=\columnwidth]{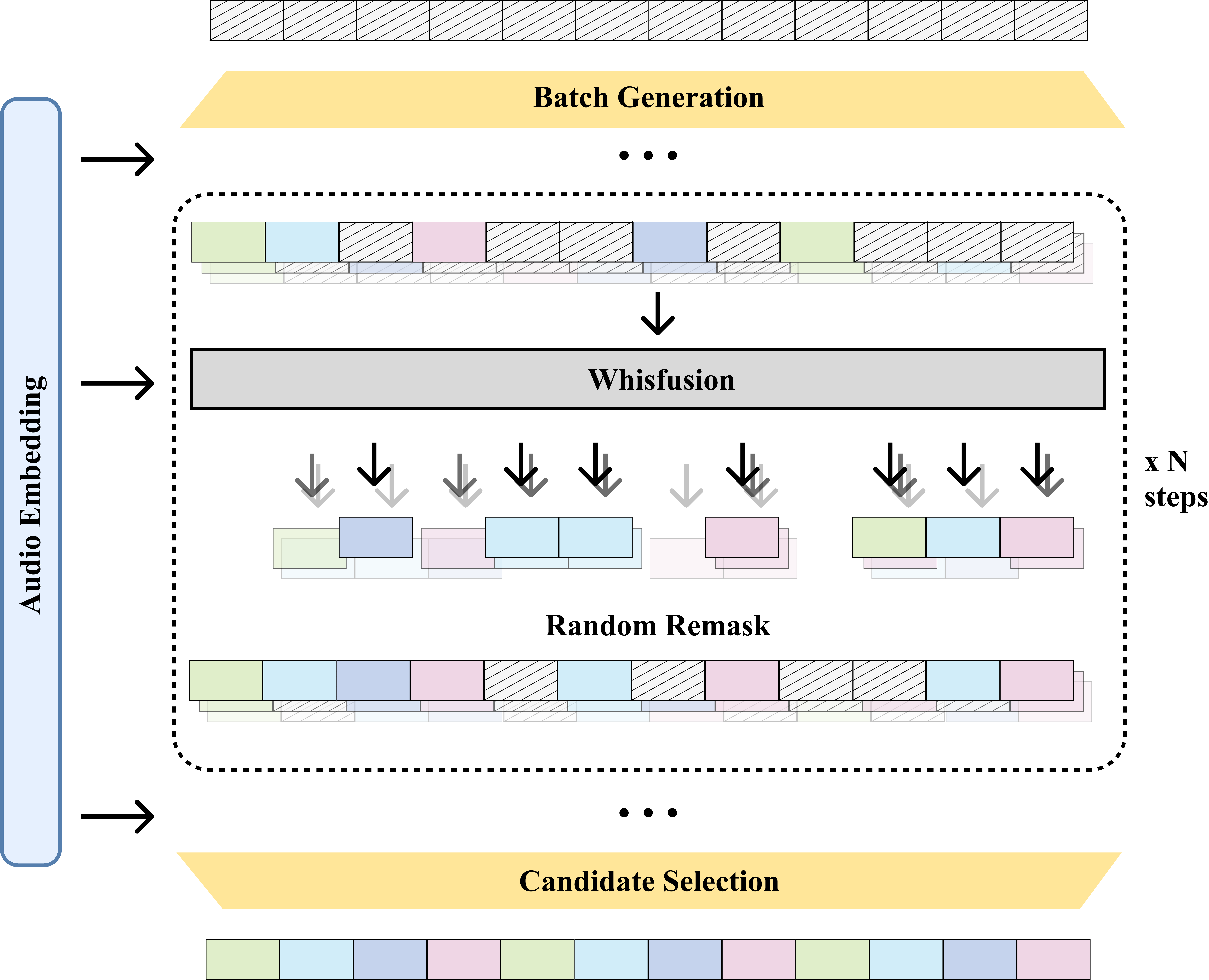}
  \caption{\textbf{Parallel Diffusion Decoding (PDD).}
  Starting from a fully masked transcript, the decoder iteratively
  denoises over a fixed number of steps with random remasking between
  steps. Multiple candidate transcripts are decoded in parallel as a
  single batch. The final hypothesis is selected by MBR consensus,
  with optional Whisper likelihood reranking.}
  \label{fig:pdd}
\end{figure}

At inference time, Whisfusion preserves the task prompt tokens and
initializes the transcript region as mask tokens. Given audio
embeddings $c$, the decoder refines the masked transcript through a
fixed number of denoising steps (Figure~\ref{fig:pdd}). At each step,
the decoder predicts all currently masked transcript positions in
parallel. We then commit
the predictions and redraw a random Bernoulli mask according to the
next target mask ratio (the final step commits all remaining masks).
We use random Bernoulli remasking rather than confidence-based
selection: the latter induces a remasking distribution that diverges
from the random masking pattern seen during training, and additionally
relies on high-mask-regime confidences that we found poorly
calibrated; both effects degraded performance in our preliminary
experiments (see Appendix~\ref{app:remask}). Our default trajectory feeds the three denoising steps
with input mask ratios $[1.0, 0.9, 0.8]$, which stays within the
high-mask regime emphasized during Stage 2 training. Thus, the sequential depth
of decoding is fixed by $N{=}3$, rather than by the transcript length.

This masked-diffusion formulation naturally enables candidate-level
parallelism: because each hypothesis is a full masked sequence, we
decode $K$ candidate transcripts as an ordinary batch on GPU. Independent sampling
and random remasking across candidates produce diverse full-sequence
hypotheses, increasing the chance that at least one candidate is close
to the target transcript. The softmax sampling temperature $\tau$
controls per-token randomness when committing predictions at each
denoising step. Increasing $K$ improves the candidate pool but also
increases decoder computation, making $K$ a speed--accuracy knob.
Unless otherwise specified, we use $K{=}5$ and $\tau{=}0.1$ as the
default high-throughput setting in the main comparisons. We also
evaluate $K{=}15$ as an accuracy-oriented setting (analyzed in
\S\ref{sec:analysis-pdd}). Sensitivity to $N$, $\tau$, and the
trajectory schedule is reported in Appendix~\ref{app:hparams}.

After PDD produces $K$ complete candidates, Whisfusion selects the
final transcript by minimum Bayes risk (MBR) \citep{kumar2004mbr}
by default. Specifically,
we choose the candidate with the minimum sum of pairwise edit
distances to the other candidates,
\begin{equation}
  \hat{y} \;=\; \arg\min_{y_i \in \mathcal{Y}}
  \sum_{j \neq i} d(y_i, y_j),
\end{equation}
where $\mathcal{Y}$ is the candidate set and $d$ is the
language-aware normalized edit rate: WER for languages with explicit
word boundaries, and CER for CJK and other languages without
reliable word boundaries (see Appendix~\ref{app:mbr} for the exact
CER-language set). This reference-free consensus criterion
uses the diversity of the PDD candidate set without requiring an
additional learned selector.

As an optional accuracy-oriented variant, we rerank the same PDD
candidates using length-normalized Whisper-large-v3 likelihood,
\begin{equation}
  \hat{y} \;=\; \arg\max_{y_i \in \mathcal{Y}}
  \frac{1}{|y_i|} \log p_{\text{Whisper}}(y_i \mid x).
\end{equation}
This does not change candidate generation, but adds an
audio-conditioned selection signal. Because Whisper reranking requires
additional decoder computation and parameters, we report it separately
from the default MBR setting and include its runtime cost in RTFx.
Algorithm~\ref{alg:pdd} in Appendix~\ref{app:algos} summarizes the
full decoding procedure.

\section{Experimental Setup}
\label{sec:setup}

\subsection{Training Data and Evaluation Benchmarks}
\label{sec:setup-data}

We train Whisfusion on approximately 68k hours of 11-language speech,
filtered to utterances of at most 30 seconds to match the Whisper
encoder window. We apply language-aware text normalization (Whisper's
English normalizer for English, basic normalizer otherwise)
consistently in training and evaluation. To mitigate the
English-heavy corpus distribution, we extend the
$\alpha$-temperature sampling of \citet{conneau2020xlmr} to a
two-level scheme with $\alpha_{\text{lang}}{=}0.3$ across languages
and $\alpha_{\text{dataset}}{=}0.5$ within each language. For Japanese
ReazonSpeech \citep{yin2023reazonspeech}, we blacklist 1k held-out
evaluation utterances from training to prevent train--test
contamination. Per-language and per-dataset hours, along with effective
sampling weights, are given in Appendix~\ref{app:data}.

We evaluate on three benchmark groups. The English group contains
five splits: LibriSpeech \citep{panayotov2015librispeech} test-clean
and test-other, Earnings-22 \citep{delrio2022earnings22}, VoxPopuli-en
\citep{wang2021voxpopuli}, and CommonVoice-en
\citep{ardila2020commonvoice}. The European group contains 20 WER
splits across seven languages (de, nl, fr, es, it, pt, pl) and three
benchmark families: MLS \citep{pratap2020mls} (7), CommonVoice (7),
and VoxPopuli (6; no Portuguese). The CJK
group contains six CER splits over Chinese, Japanese, and Korean:
CV-zh, AISHELL-zh \citep{bu2017aishell}, CV-ja, Reazon-ja, FLEURS-ja
\citep{conneau2022fleurs}, and Kspon-ko \citep{bang2020kspon}.
CommonVoice splits, Reazon-ja, and Kspon-ko use a fixed 1k random
subsample (seed 42); other splits use the full 30s-filtered test set.
Unless otherwise specified, reported averages are unweighted
macro-averages over evaluation splits. The complete list of evaluation
splits and subsampling seeds is given in Appendix~\ref{app:splits},
and the exact per-language text normalizers used in both training and
evaluation are summarized in Appendix~\ref{app:normalize}.

\subsection{Baselines}
\label{sec:setup-baselines}

We compare Whisfusion against four groups of ASR systems. For
autoregressive baselines, we include Whisper-large-v3, Whisper-turbo,
Canary-1b-v2, and Qwen3-ASR-1.7B, which represent strong modern
encoder-decoder or audio-language ASR systems. For fast
non-autoregressive baselines, we include OWSM-CTC v3.1
\citep{peng2024owsmctc} and MMS-all \citep{pratap2023mms} as
CTC-based systems. For diffusion-style or diffusion-LLM ASR, we
include Drax and Whisper-LLaDA as publicly evaluable baselines. For
Whisper-LLaDA, we use the standalone (non-cascade) decoding mode for
a fair comparison with single-model NAR systems.

MDM-ASR and dLLM-ASR lack public checkpoints and cannot be
evaluated under our unified protocol; we therefore exclude them
from our speed--accuracy comparisons. For baselines with missing splits (e.g., Drax on some
multilingual benchmarks), we average over available splits and mark
them accordingly. Exact checkpoints, decoding hyperparameters, and
audio preprocessing for all baselines are listed in
Appendix~\ref{app:baselines}.

\subsection{Metrics and Runtime Protocol}
\label{sec:setup-metrics}

We report WER for English/European benchmarks and CER for CJK, where
word boundaries are unreliable. Throughout the paper, Whisfusion
denotes the default setting (\S\ref{sec:method-pdd}) and Whisfusion +
rerank the reranked variant.

We report runtime using RTFx, where higher is faster. Formally,
\begin{equation}
  \mathrm{RTFx} \;=\;
  \frac{\sum_i \mathrm{duration}(x_i)}{\sum_i \mathrm{time}(x_i)}.
\end{equation}
Timing covers encoder forward, decoding, candidate selection, and
Whisper reranking (when used); file I/O is excluded. For fair
comparison across systems, all runtime numbers
are measured on a single H100 with batch size 1, bf16 precision, and
no compilation, following the convention used by recent generative
NAR ASR baselines \citep{navon2025drax,yen2026mdmasr}. RTFx is
measured on a length-balanced 336-utterance English subset,
while WER/CER are computed on the full evaluation splits described
in \S\ref{sec:setup-data}. Full measurement protocol details are
described in Appendix~\ref{app:rtfx}.

\begin{table*}[!t]
  \centering
  \small
  \begin{tabular}{clcccccccc}
  \toprule
   & \multirow{2}{*}{\textbf{Model}}
   & \multicolumn{6}{c}{\textbf{WER} (\%) $\downarrow$}
   & \multirow{2}{*}{\textbf{Params (B)}}
   & \multirow{2}{*}{\textbf{RTFx} $\uparrow$} \\
   & & LS-clean & LS-other & Earnings-22 & VoxPopuli & CV-en & Avg & & \\
  \midrule
  \multirow{4}{*}{\rotatebox[origin=c]{90}{\textbf{AR}}}
   & Whisper-large-v3      & 1.97 & 4.31 & 10.64 & 8.96 & 9.65 & 7.11 & 1.6 & 35.33 \\
   & Whisper-turbo         & 2.10 & 4.05 & 10.88 & 11.24 & 16.03 & 8.86 & 0.8 & 143.99 \\
   & Canary-1b-v2          & 2.09 & 3.68 & 11.36 & \textbf{5.88} & \underline{8.97} & 6.40 & 1.0 & 47.87 \\
   & Qwen3-ASR-1.7B        & \underline{1.69} & \textbf{3.49} & \textbf{9.32} & \underline{5.97} & \textbf{7.61} & \textbf{5.62} & 2.0 & 20.12 \\
  \midrule
  \multirow{6}{*}{\rotatebox[origin=c]{90}{\textbf{NAR}}}
   & OWSM-CTC v3.1         & 2.46 & 5.30 & 12.71 & 8.13 & 12.77 & 8.27 & 1.0 & 665.74 \\
   & MMS-all               & 3.74 & 8.08 & 22.78 & 8.53 & 21.50 & 12.93 & 1.0 & 681.21 \\
  \cmidrule(lr){2-10}
   & Whisper-LLaDA         & 2.41 & 4.88 & 16.11 & 10.47 & 15.22 & 9.82 & 8.7 & 7.92 \\
   & Drax (MBR, 8/16)      & 2.40 & 5.34 & 14.09 & 7.07 & 12.78 & 8.34 & 1.2 & 33.16 \\
  \cmidrule(lr){2-10} \\[-1.4em] \cmidrule(lr){2-10}
   & Whisfusion (ours)     & 1.74 & 3.78 & 10.33 & 6.66 & 10.25 & 6.55 & 1.5 & 173.45 \\
   & Whisfusion + rerank   & \textbf{1.67} & \underline{3.53} & \underline{9.64} & 6.34 & 9.15 & \underline{6.07} & 2.4 & 143.27 \\
  \bottomrule
  \end{tabular}
  \caption{\textbf{English ASR speed--accuracy comparison.}
  WER (\%) on five English benchmarks; Avg is the macro-average over
  the five splits. RTFx is measured on a single H100
  (\S\ref{sec:setup-metrics}).}
  \label{tab:en-main}
\end{table*}

\section{Main Results}
\label{sec:main}

\subsection{English Speed--Accuracy Trade-off}
\label{sec:main-en}

Table~\ref{tab:en-main} compares Whisfusion with strong English ASR
baselines across five benchmarks. The default Whisfusion setting
establishes a high-throughput NAR operating point: it achieves 6.55
average WER at 173 RTFx, outperforming Whisper-turbo in both accuracy
and throughput. Compared with Whisper-large-v3, Whisfusion reduces
the average WER from 7.11 to 6.55 while running nearly five times
faster. With optional Whisper reranking, Whisfusion further reduces
the average WER to 6.07 while maintaining 143 RTFx, approximately
matching Whisper-turbo-level throughput with substantially lower
error.

The table also shows that Whisfusion occupies a different point from
both CTC-style NAR systems and diffusion/dLLM baselines. OWSM-CTC and
MMS-all achieve very high throughput, but their average WERs remain
substantially higher than Whisfusion's. In contrast, public diffusion
and dLLM baselines such as Drax and Whisper-LLaDA are both less
accurate and slower than Whisfusion under our protocol. Against
stronger AR systems, Whisfusion trades a small amount of accuracy for
much higher throughput: the default model is within $0.15$\,pp of
Canary-1b-v2 while running 3.6$\times$ faster. With Whisper reranking,
Whisfusion surpasses Canary and approaches Qwen3-ASR within $0.45$\,pp
at approximately Whisper-turbo-level throughput.

\subsection{European Multilingual ASR}
\label{sec:main-eu}

\begin{table}[t]
  \centering
  \small
  \setlength{\tabcolsep}{3pt}
  \begin{tabular}{lcccc}
  \toprule
  \multirow{2}{*}{\textbf{Model}}
   & \multicolumn{4}{c}{\textbf{WER} (\%) $\downarrow$} \\
   & MLS (7) & CV (7) & VP (6) & All (20) \\
  \midrule
  Whisper-large-v3     & \textbf{6.52}    & 6.23  & 16.01 & 9.27 \\
  Whisper-turbo        & \underline{6.62} & 7.31  & 19.46 & 10.71 \\
  Canary-1b-v2         & 7.09             & 6.29  & \textbf{9.82}  & \textbf{7.63} \\
  Qwen3-ASR-1.7B       & 9.18             & 7.15  & 12.24 & 9.39 \\
  \cmidrule(lr){1-5}
  OWSM-CTC v3.1        & 19.12 & 18.40 & 21.21 & 19.50 \\
  MMS-all              & 10.57 & 11.48 & 12.65 & 11.51 \\
  \cmidrule(lr){1-5}
  Drax (MBR 8/16)      & 8.16  & 9.08  & 11.59 & 9.47 \\
  \cmidrule(lr){1-5} \\[-1.4em] \cmidrule(lr){1-5}
  Whisfusion (ours)    & 8.81  & \underline{4.94} & 11.27 & 8.19 \\
  Whisfusion + rerank  & 8.22  & \textbf{4.40}    & \underline{10.93} & \underline{7.70} \\
  \bottomrule
  \end{tabular}
  \caption{\textbf{European WER by benchmark family.}
  WER (\%) averaged within MLS, CommonVoice (CV), and VoxPopuli (VP);
  All is the macro-average over all 20 European splits. Drax is
  averaged over the 14 of 20 splits where it provides outputs;
  on those splits, Whisfusion obtains 8.14 WER (7.64 with rerank)
  versus Drax 9.47.}
  \label{tab:eu-main}
\end{table}

\begin{table*}[!t]
  \centering
  \small
  \begin{tabular}{lccccccc}
  \toprule
  \multirow{2}{*}{\textbf{Model}}
   & \multicolumn{2}{c}{\textbf{ZH}}
   & \multicolumn{3}{c}{\textbf{JA}}
   & \textbf{KO}
   & \multirow{2}{*}{\textbf{Avg (6)}} \\
  \cmidrule(lr){2-3} \cmidrule(lr){4-6} \cmidrule(lr){7-7}
   & CV-zh & AISHELL-zh & CV-ja & Reazon-ja & FLEURS-ja & Kspon-ko & \\
  \midrule
  Whisper-large-v3     & 15.83 & 9.25 & 22.40 & 15.23 & 6.14 & 13.76 & 13.77 \\
  Whisper-turbo        & 14.50 & 9.20 & 26.24 & 11.27 & \underline{6.08} & 13.88 & 13.53 \\
  Qwen3-ASR-1.7B       & \textbf{5.80} & \textbf{1.54} & 21.17 & 26.41 & \textbf{5.45} & 9.64 & 11.67 \\
  \cmidrule(lr){1-8}
  OWSM-CTC v3.1        & \underline{12.75} & 6.38 & 22.74 & \textbf{10.03} & 7.69 & 15.66 & 12.54 \\
  MMS-all              & 25.59 & 31.13 & 39.92 & 50.37 & 20.98 & 41.26 & 34.88 \\
  \cmidrule(lr){1-8}
  Drax (MBR 8/16)      & 16.68 & 7.13 & 22.78 & 11.45 & 9.74 & --- & 13.56 \\
  \cmidrule(lr){1-8} \\[-1.4em] \cmidrule(lr){1-8}
  Whisfusion (ours)    & 15.98 & 5.55 & \textbf{12.69} & 10.27 & 12.89 & \underline{9.09} & \underline{11.08} \\
  Whisfusion + rerank  & 14.49 & \underline{4.93} & \underline{12.72} & \underline{10.04} & 12.39 & \textbf{8.34} & \textbf{10.48} \\
  \bottomrule
  \end{tabular}
  \caption{\textbf{CJK ASR results.} CER (\%) on six Chinese, Japanese,
  and Korean splits; Avg is the macro-average over the six splits.
  Drax is averaged over the 5 of 6 splits where it provides outputs
  (no Kspon-ko output); on the same 5-split overlap, Whisfusion
  obtains 11.48 CER and Whisfusion + rerank 10.91 versus Drax 13.56.}
  \label{tab:cjk-main}
\end{table*}

Table~\ref{tab:eu-main} evaluates whether Whisfusion's gains transfer
beyond English to multilingual European ASR. Under the same default
$K{=}5$ setting used in Table~\ref{tab:en-main}, Whisfusion achieves
8.19 average WER over 20 European splits, outperforming
Whisper-large-v3, Qwen3-ASR, and CTC-based NAR baselines while
remaining within 0.56\,pp of Canary-1b-v2 and comparing favorably
to Drax on its available splits. With Whisper reranking, the
average drops to 7.70, essentially matching Canary-1b-v2 (within
0.07\,pp) and confirming multilingual transfer.

The gains are particularly pronounced on CommonVoice (4.94 WER,
4.40 with rerank), substantially outperforming all baselines.
At the same time, the results are not uniform across domains:
Whisper-large-v3 remains strongest on MLS, while Canary-1b-v2 is
strongest on VoxPopuli, suggesting Whisfusion is a strong overall
operating point (per-split numbers in
Appendix~\ref{app:per-split}).

\subsection{CJK Transcription Results}
\label{sec:main-cjk}

Table~\ref{tab:cjk-main} reports CER on CJK benchmarks, where
word-level evaluation is unreliable. Whisfusion achieves the best
average CER among evaluated systems: the default model obtains 11.08
average CER, and Whisper reranking further improves it to 10.48. This
indicates that the speech-conditioned masked diffusion decoder
transfers beyond space-delimited languages and remains competitive in
character-level transcription settings.

The gains, however, are uneven across languages and datasets.
Whisfusion is strongest on CV-ja and Kspon-ko, and remains highly
competitive on Reazon-ja. In contrast, Qwen3-ASR is substantially
stronger on Chinese, with much lower CER on both CV-zh and AISHELL-zh,
and also performs strongly on FLEURS-ja. We therefore interpret the
CJK result as evidence of strong average multilingual performance,
while identifying Chinese and some Japanese evaluation settings as
areas for further improvement.

\section{Analysis and Ablations}
\label{sec:analysis}

The results above show that Whisfusion reaches a strong
speed--accuracy operating point. We next analyze two components
central to this behavior. First, we test whether high-mask
specialization improves train--inference alignment for fully masked
decoding. Second, we study how PDD candidate generation, candidate
selection, and candidate set size jointly determine the
speed--accuracy trade-off.

\subsection{High-Mask Specialization}
\label{sec:analysis-highmask}

\begin{table}[t]
  \centering
  \small
  \setlength{\tabcolsep}{4pt}
  \begin{tabular}{lcccccc}
  \toprule
  \multirow{2}{*}{\textbf{Stage~2}}
   & \multicolumn{3}{c}{\textbf{EN (5)}}
   & \multicolumn{3}{c}{\textbf{EU (20)}} \\
  \cmidrule(lr){2-4} \cmidrule(lr){5-7}
   & $N{=}1$ & $N{=}2$ & $N{=}3$ & $N{=}1$ & $N{=}2$ & $N{=}3$ \\
  \midrule
  Uniform                 & 14.3 & 7.5 & 6.8 & 23.7 & 9.7 & 8.4 \\
  High-mask               & \textbf{12.5} & \textbf{7.2} & \textbf{6.6}
                          & \textbf{20.6} & \textbf{9.6} & \textbf{8.2} \\
  \cmidrule(lr){1-7}
  $\Delta$\,MBR           & $-1.8$ & $-0.3$ & $-0.2$ & $-3.1$ & $-0.1$ & $-0.2$ \\
  $\Delta$\,Oracle        & $-1.7$ & $-0.1$ & $-0.3$ & $-2.8$ & $-0.1$ & $-0.2$ \\
  \bottomrule
  \end{tabular}
  \caption{\textbf{High-mask specialization vs.\ uniform mask control.}
  MBR WER (\%) on EN (5) / EU (20) under $N\in\{1,2,3\}$ denoising
  steps ($K{=}5$, $\tau{=}0.1$, default trajectory truncated to the
  first $N$). Stage~2 variants share the Stage~1 checkpoint and
  training budget; only the mask-ratio distribution differs
  (uniform: $U(0,1)$; high-mask: $U(0.7,1.0)$). $\Delta$ rows:
  high-mask minus uniform on MBR and oracle@$K{=}5$.}
  \label{tab:highmask}
\end{table}

We first test whether Stage~2's gains come from additional
optimization or specifically from narrowing the mask distribution.
We resume the same Stage~1 checkpoint and continue training for an
identical 10-epoch budget under two regimes: a uniform control that
keeps $t \sim U(0,1)$ and our recipe that narrows to
$t \sim U(0.7, 1.0)$, holding all other training conditions fixed
so that only the per-step mask-ratio distribution differs.

Table~\ref{tab:highmask} reports MBR WER under three inference step
counts $N \in \{1, 2, 3\}$. At $N{=}1$, a single denoising step
operates entirely on a fully masked input; high-mask specialization
improves over the uniform control by 1.8~pp on English and 3.1~pp
on European. As $N$ grows, iterative refinement progressively
compensates for a weaker first step in the uniform model: the gap
shrinks to 0.3/0.1~pp at $N{=}2$ and 0.2/0.2~pp at $N{=}3$. This is
the behavior predicted by our training argument in
\S\ref{sec:method-train}: when no iterative correction is available,
the prediction quality at $t{=}1.0$ is the dominant factor, which is
precisely what high-mask training targets.

The $\Delta$\,Oracle row tracks $\Delta$\,MBR within 0.3~pp at every
cell (and within 0.2~pp at $N{\geq}2$), indicating that the gains
come from higher-quality candidate generation rather than from
candidate selection:
even an oracle that picks the best of the $K{=}5$ candidates benefits
by essentially the same margin as MBR. We interpret this as evidence that high-mask
specialization improves the underlying PDD trajectory itself, with
MBR preserving that improvement at selection time. While the gain at
$N{=}3$ is small, the substantial $N{=}1$ improvement
($-1.8$/$-3.1$\,pp) shows that high-mask specialization is what
makes very-low-step inference viable: it decouples the
\emph{training} mask distribution from the \emph{inference} step
count, giving Whisfusion a smooth speed--accuracy knob beyond the
default $N{=}3$ setting.

\subsection{PDD Candidate Selection and Operating Points}
\label{sec:analysis-pdd}

\begin{figure}[t]
  \centering
  \includegraphics[width=\columnwidth]{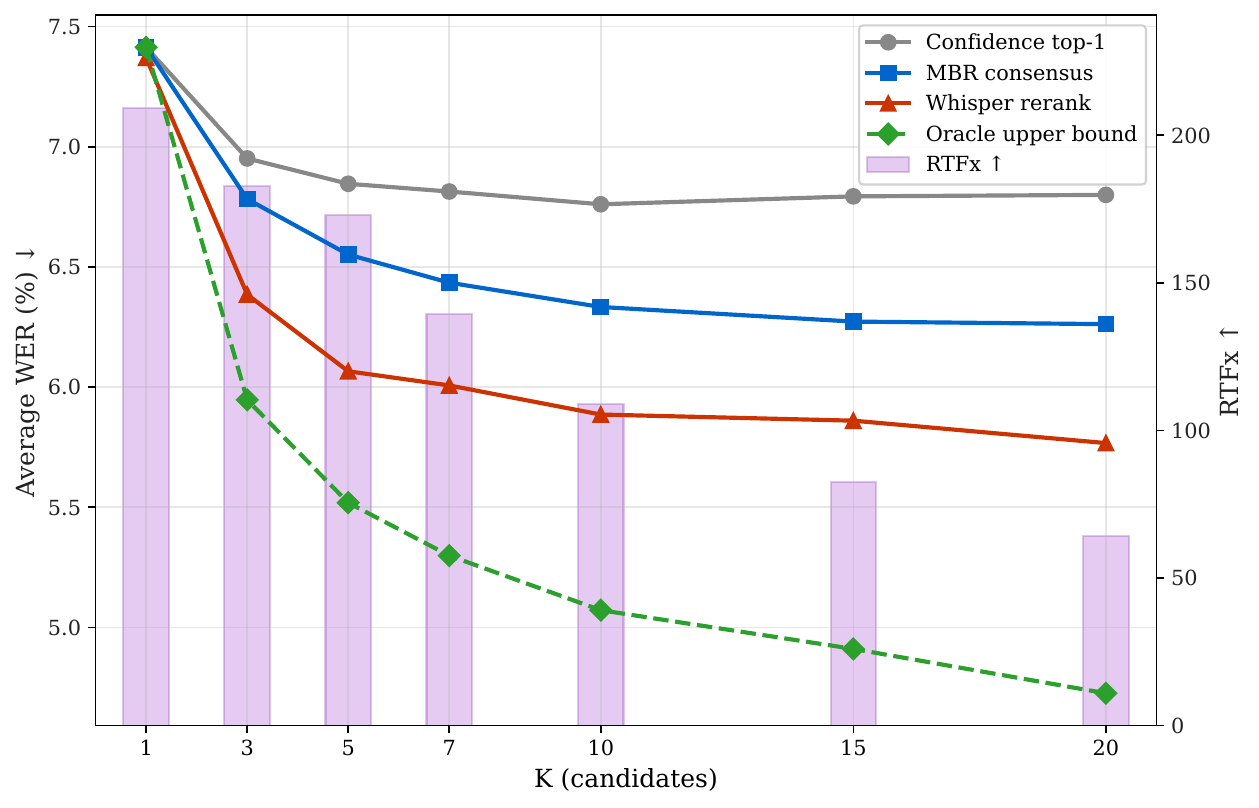}
  \caption{\textbf{PDD candidate selection and throughput as $K$
  varies.} Lines show average WER over the five English splits for
  confidence top-1, MBR, Whisper reranking, and oracle selection;
  bars show RTFx for MBR selection (\S\ref{sec:setup-metrics}).}
  \label{fig:k-sweep}
\end{figure}

Figure~\ref{fig:k-sweep} analyzes how the number of PDD candidates
affects both candidate quality and throughput. As $K$ increases, the
oracle WER decreases sharply, indicating that larger candidate sets
contain increasingly accurate hypotheses. However, confidence top-1
selection improves much less, showing that the model often generates
better candidates than it selects. MBR consistently improves over
confidence selection by using agreement among candidates as a
reference-free signal, and Whisper reranking further reduces WER by
adding an audio-conditioned likelihood score without changing
candidate generation. The remaining gap between Whisper reranking
and the oracle upper bound suggests that PDD candidate generation is
strong, but candidate selection remains an important source of
headroom.

Increasing $K$ incurs a throughput cost. Although PDD batches candidates and
therefore benefits from GPU parallelism, decoder computation still
grows with the candidate batch size, and MBR scoring adds a quadratic
pairwise-distance cost. As shown by the RTFx bars, throughput
decreases as $K$ grows. We therefore use $K{=}5$ as the default
high-throughput setting in the main comparisons, where it already
provides a strong speed--accuracy point. Larger candidate sets such
as $K{=}15$ serve as accuracy-oriented settings: they further reduce
WER, but with diminishing gains. Per-setting numbers and additional
sensitivity sweeps for the trajectory schedule, step count $N$, and
sampling temperature $\tau$ are provided in
Appendix~\ref{app:hparams}.

\section{Conclusion}
\label{sec:conclusion}

We presented Whisfusion, a speech-conditioned masked diffusion
framework that grounds a from-scratch decoder in frozen
Whisper-large-v3 audio embeddings and refines transcripts in a
fixed number of denoising steps, replacing left-to-right token
generation. Our results show that non-autoregressive ASR can move
beyond the conventional trade-off between fast alignment-based
decoding and high-quality autoregressive decoding: with large-scale
multilingual training, high-mask specialization, and consensus-based
PDD, Whisfusion reaches the accuracy regime of modern autoregressive
ASR systems while retaining the high-throughput advantage of parallel
inference. These findings establish masked diffusion as a
Pareto-competitive paradigm for scalable, high-throughput
multilingual ASR.

\section*{Limitations}

First, Whisfusion currently inherits the scope of its frozen
Whisper-large-v3 encoder: we evaluate utterances up to 30 seconds,
and extending the approach to long-form or streaming ASR would
require additional chunking or memory reuse. Second, our oracle analysis
indicates that candidate selection remains a bottleneck: PDD can
generate better hypotheses than those selected by MBR or Whisper
reranking, leaving room for learned rerankers or audio-conditioned
risk optimization. Third, our experiments focus on transcription and
do not evaluate speech-language tasks such as audio question
answering, instruction following, or long-form generation.


\bibliography{anthology,custom}

\appendix

\section{Implementation Details}
\label{app:impl}

\subsection{Decoder Architecture}
\label{app:arch}

Table~\ref{tab:arch} lists the full Whisfusion decoder
specification. The decoder follows a LLaMA-style transformer stack
with cross-attention to the frozen Whisper-large-v3 audio embeddings
\citep{radford2023whisper} in every block.

\begin{table}[h]
\centering
\footnotesize
\setlength{\tabcolsep}{3pt}
\begin{tabular}{lc}
\toprule
\textbf{Component} & \textbf{Value} \\
\midrule
\multicolumn{2}{l}{\emph{Decoder shape}} \\
Layers $/$ Hidden $/$ Heads & $24 \,/\, 1280 \,/\, 20$ \\
Head dim & $64$ \\
FFN intermediate (SwiGLU) & $4096$ \\
Max text sequence length & $192$ \\
\midrule
\multicolumn{2}{l}{\emph{Normalization and position}} \\
Norm & RMSNorm, $\varepsilon{=}10^{-5}$ \\
QK-normalization & enabled \\
Position encoding & RoPE, base $10^{4}$, full \\
Bias / embedding tying & none / untied \\
\midrule
\multicolumn{2}{l}{\emph{Vocabulary}} \\
Tokenizer & Whisper-large-v3 \\
Vocab size (base $+$ mask) & $51866 + 1$ \\
Prompt prefix preserved & $4$ tokens \\
\midrule
\multicolumn{2}{l}{\emph{Cross-attention}} \\
Cross-attention & every block \\
Audio projection & Linear-GELU-Linear, $1280$ \\
Audio cache (inference) & built once per utterance \\
\midrule
\multicolumn{2}{l}{\emph{Parameters}} \\
Decoder (trainable) & $828$\,M \\
Whisper encoder (frozen) & $635$\,M \\
\bottomrule
\end{tabular}
\caption{\textbf{Whisfusion decoder architecture.}
The decoder is trained from scratch; the Whisper-large-v3
encoder is kept frozen throughout training and inference.}
\label{tab:arch}
\end{table}

We initialize all linear and embedding weights with
$\mathcal{N}(0, 0.02^{2})$ and rescale the self-attention,
cross-attention, and SwiGLU down-projection output weights by
$1 / \sqrt{2N}$, where $N{=}24$ is the layer count, following the
GPT-NeoX/OLMo deep-model stability recipe. The added mask token
uses the same normal initialization. Cross-attention queries come
from the decoder hidden states while keys and values come from the
audio embeddings; no position encoding is applied on the audio
side. At inference, the audio key/value tensors are computed once
per utterance and reused across all $N$ denoising steps and $K$
PDD candidates.

\subsection{Training Schedule}
\label{app:train}

Table~\ref{tab:train} summarizes the two-stage training schedule.
Both stages use the same data, sampler, and architecture; they
differ only in the mask sampling range, peak learning rate,
warmup length, and number of epochs. Stage 2 is initialized from
the Stage 1 exponential-moving-average (EMA) checkpoint at
epoch~12, with the optimizer state and learning-rate schedule
restarted at the Stage 2 peak.

\begin{table}[h]
\centering
\footnotesize
\setlength{\tabcolsep}{2pt}
\begin{tabular}{lcc}
\toprule
\textbf{Setting} & \textbf{Stage 1} & \textbf{Stage 2} \\
\midrule
Mask range $t$
   & $U(0, 1)$ & $U(0.7, 1.0)$ \\
Initialization
   & random & S1 ep.\ 12 \\
Optimizer
   & \multicolumn{2}{c}{AdamW, $(\beta_1, \beta_2) = (0.9, 0.95)$} \\
Weight decay
   & \multicolumn{2}{c}{$0.1$} \\
Peak learning rate
   & $2{\times}10^{-4}$ & $6{\times}10^{-5}$ \\
LR schedule
   & \multicolumn{2}{c}{cosine to $10\%$ of peak} \\
Warmup (linear)
   & $2{,}000$ steps & $1{,}000$ steps \\
Gradient clip ($L_{2}$)
   & \multicolumn{2}{c}{$1.0$} \\
Epochs
   & $12$ & $10$ \\
Total optimizer steps
   & ${\sim}134$k & $\sim 112$k \\
Batch (per GPU $/$ global)
   & \multicolumn{2}{c}{$64\,/\,2{,}048$ utterances} \\
Max text length
   & \multicolumn{2}{c}{$192$ tokens} \\
Precision
   & \multicolumn{2}{c}{bf16-mixed} \\
EMA decay
   & \multicolumn{2}{c}{$0.999$} \\
Frozen modules
   & \multicolumn{2}{c}{Whisper-large-v3 encoder} \\
\bottomrule
\end{tabular}
\caption{\textbf{Two-stage training schedule.}}
\label{tab:train}
\end{table}

The training objective is cross-entropy over masked positions only;
padding tokens and the four-token prompt prefix are excluded from
the loss (Appendix~\ref{app:objective}). All training uses 4 nodes
of 8 NVIDIA H100 80\,GB GPUs (32 GPUs total) under PyTorch
Lightning Fabric with FSDP (\textsc{full\_shard}); decoder weights
are compiled with \texttt{torch.compile} and activation
checkpointing is enabled per decoder block. Wall-clock time was
approximately $38$\,h for Stage 1 (through epoch~12) and $32$\,h
for Stage 2, corresponding to about $1{,}216$ and $1{,}024$
H100-hours respectively.

\subsection{Training Data Composition}
\label{app:data}

Whisfusion is trained on approximately $68$k hours of $11$-language
speech ($22.9$M utterances) distributed over $40$
(language $\times$ dataset) buckets drawn from public corpora.
Utterances are filtered to between $1$ and $30$ seconds to match
the Whisper encoder window and to drop short backchannels. To
align the training distribution with our evaluation domain mix,
all English splits inside the multilingual manifest are dropped
and English is re-included only from CommonVoice, VoxPopuli, and
Earnings-22. A fixed $1{,}000$-utterance ReazonSpeech-ja audio
subset (seed $42$) is blacklisted from training to prevent
contamination with the held-out evaluation split
(Appendix~\ref{app:splits}).

\begin{table}[h]
\centering
\small
\begin{tabular}{lrrrr}
\toprule
\textbf{Lang} & \textbf{Hours} & \textbf{Nat.\,\%}
              & \textbf{Tgt.\,\%} & \textbf{Boost} \\
\midrule
en & $51{,}099$ & $74.8$ & $23.2$ & $0.31\times$ \\
ja &  $5{,}318$ &  $7.8$ & $11.8$ & $1.51\times$ \\
de &  $3{,}557$ &  $5.2$ & $10.4$ & $2.00\times$ \\
fr &  $2{,}310$ &  $3.4$ &  $9.2$ & $2.71\times$ \\
nl &  $1{,}700$ &  $2.5$ &  $8.4$ & $3.35\times$ \\
es &  $1{,}631$ &  $2.4$ &  $8.3$ & $3.45\times$ \\
ko &     $970$ &  $1.4$ &  $7.1$ & $4.97\times$ \\
it &     $636$ &  $0.9$ &  $6.2$ & $6.69\times$ \\
zh &     $381$ &  $0.6$ &  $5.3$ & $9.52\times$ \\
pl &     $354$ &  $0.5$ &  $5.2$ & $10.0\times$ \\
pt &     $326$ &  $0.5$ &  $5.1$ & $10.6\times$ \\
\bottomrule
\end{tabular}
\caption{\textbf{Per-language training hours and effective
sampling weights.} \textbf{Nat.\,\%} is the natural share of
training hours; \textbf{Tgt.\,\%} is the marginal share after
two-level $\alpha$-temperature sampling with
$\alpha_{\text{lang}}{=}0.3$ and
$\alpha_{\text{dataset}}{=}0.5$; \textbf{Boost} is target divided
by natural share.}
\label{tab:data}
\end{table}

We use two-level $\alpha$-temperature sampling
\citep{conneau2020xlmr} extended over languages and datasets:
language $\ell$ is sampled with $p(\ell) \propto h(\ell)^{0.3}$,
where $h(\ell)$ is the total hours for $\ell$, and dataset $d$
within $\ell$ with $p(d \mid \ell) \propto h(\ell, d)^{0.5}$.
Per-utterance weights are normalized by the bucket utterance
count so that the marginal sampling distribution matches the
target shares in Table~\ref{tab:data} up to sampling variance,
independent of per-utterance duration.

\subsection{Algorithms}
\label{app:algos}

Algorithm~\ref{alg:training} summarizes the two-stage training
procedure of \S\ref{sec:method-train}, and
Algorithm~\ref{alg:pdd} summarizes the Parallel Diffusion Decoding
procedure of \S\ref{sec:method-pdd}.

\begin{algorithm}[t]
\caption{Two-Stage Training for Whisfusion.
Per-stage hyperparameters are in Table~\ref{tab:train}.}
\label{alg:training}
\begin{algorithmic}[1]
\Require Frozen Whisper encoder $E_\phi$; MDM decoder $D_\theta$;
         dataset $\mathcal{D}$; prompt-preserve mask $m$
\Ensure  EMA-averaged decoder parameters $\theta_{\text{EMA}}$
\State Initialize $\theta$ randomly; $\theta_{\text{EMA}} \gets \theta$
\State \textbf{--- Stage 1: Uniform mask training ---}
\For{epoch $= 1$ to $N_1$}
  \For{batch $(x, y_0) \in \mathcal{D}$}
    \State $c \gets E_\phi(x)$
    \State $t \sim U(0, 1)$
    \State $y_t \gets \textsc{Mask}(y_0, t, m)$
       \Comment{prompt prefix kept intact}
    \State $\mathcal{L} \gets -\!\!\sum_{i \in \mathcal{M}}
           \log p_\theta(y_{0,i} \mid y_t, c)$
       \Comment{$\mathcal{M}$: masked, non-prompt, non-pad}
    \State $\theta \gets \textsc{AdamWStep}(\theta, \nabla\mathcal{L})$
    \State $\theta_{\text{EMA}} \gets 0.999\,\theta_{\text{EMA}}
           + 0.001\,\theta$
  \EndFor
\EndFor
\State \textbf{--- Stage 2: High-mask specialization ---}
\State $(\theta, \theta_{\text{EMA}}) \gets$ Stage 1 epoch-$12$
       checkpoint
\State Reset AdamW state; restart LR schedule at the Stage 2 peak
\For{epoch $= 1$ to $N_2$}
  \For{batch $(x, y_0) \in \mathcal{D}$}
    \State $c \gets E_\phi(x)$
    \State $t \sim U(0.7, 1.0)$
    \State $y_t \gets \textsc{Mask}(y_0, t, m)$
    \State Update $(\theta, \theta_{\text{EMA}})$ as in Stage 1
  \EndFor
\EndFor
\State \Return $\theta_{\text{EMA}}$
\end{algorithmic}
\end{algorithm}

\begin{algorithm}[t]
\caption{Parallel Diffusion Decoding (PDD)}
\label{alg:pdd}
\begin{algorithmic}[1]
\Require Audio $x$; trained Whisfusion model $M$; candidates $K$;
         steps $N$; mask-ratio trajectory $(r_1, \ldots, r_N)$;
         temperature $\tau$; selection mode $s \in \{\text{MBR},
         \text{rerank}\}$
\Ensure  Predicted transcript $\hat{y}$
\State $c \gets E_\phi(x)$
\State $Y \gets K$ copies of $[\,p_{1:P},\,
       \langle\textsc{mask}\rangle^{L - P}\,]$
\For{$n = 1$ to $N$}
  \State $\textit{logits} \gets M(Y, c)$
  \State $Y \gets \textsc{Sample}(Y, \textit{logits}, \tau)$
  \If{$n < N$}
    \State $Y \gets \textsc{RandomRemask}(Y, r_{n+1})$
  \EndIf
\EndFor
\If{$s = \text{MBR}$}
  \State $\hat{y} \gets \arg\min_i \sum_{j \neq i} d(y_i, y_j)$
\Else
  \State $\hat{y} \gets \arg\max_i \tfrac{1}{|y_i|}
         \log p_{\text{Whisper}}(y_i \mid x)$
\EndIf
\State \Return $\hat{y}$
\end{algorithmic}
\end{algorithm}

\subsection{Training Objective}
\label{app:objective}

We optimize a masked cross-entropy loss averaged over the
corrupted, non-prompt, non-pad transcript positions. Given an
audio input $x$, its transcript $y_0$, the prompt-preserve mask
$m$, and a sampled mask ratio $t$, we form $y_t$ by independently
replacing each non-preserved position with
$\langle\textsc{mask}\rangle$ with probability $t$. Let
\begin{equation*}
\mathcal{M} = \{\, i : y_{t,i} = \langle\textsc{mask}\rangle,\;
m_i = 0,\; y_{0,i} \neq \textsc{pad}\,\}
\end{equation*}
be the set of positions contributing to the loss. The training
objective is
{\small
\begin{equation}
\mathcal{L}(\theta) =
\mathbb{E}_{x, y_0, t}\!\left[
\frac{1}{|\mathcal{M}|} \sum_{i \in \mathcal{M}}
-\log p_\theta(y_{0,i} \mid y_t, c)
\right],
\end{equation}
}
where $c = E_\phi(x)$ are the frozen Whisper-large-v3 audio
embeddings. We use uniform weighting across mask ratios rather
than the $1/t$ scaling sometimes seen in masked-diffusion training,
following recent masked-diffusion language models
\citep{sahoo2024mdlm, ou2024radd} and concurrent NAR ASR
diffusion work \citep{yen2026mdmasr}. This avoids the optimization
instability of $1/t$ weights as $t \to 0$ while remaining a valid
masked-diffusion objective under the absorbing-state ELBO.

\section{Evaluation Protocol}
\label{app:eval}

\subsection{Benchmark Splits}
\label{app:splits}

Whisfusion is evaluated on $31$ splits: $5$ English (WER), $20$
European (WER) across $7$ languages and $3$ benchmark families, and
$6$ CJK (CER) across $3$ languages. CommonVoice splits, Reazon-ja,
and Kspon-ko use a fixed $1{,}000$-utterance random subsample with
\texttt{seed=42}; the remaining splits use the full official test
set, restricted to utterances at most $30$ seconds long to match
the Whisper encoder window. For Earnings-22 we use the ESB test
partition (the same six source ids used by the Open ASR
Leaderboard \citep{srivastav2023openasrleaderboard}). Korean
coverage relies on KsponSpeech alone; CommonVoice-ko is not part
of the evaluation set due to its very small native test partition.

\begin{table}[h]
\centering
\small
\setlength{\tabcolsep}{6pt}
\begin{tabular}{llr}
\toprule
\textbf{Group} & \textbf{Split} & $n$ \\
\midrule
\multirow{5}{*}{EN}
  & LS test-clean        & $2{,}611$ \\
  & LS test-other        & $2{,}932$ \\
  & Earnings-22 (ESB)    & $2{,}704$ \\
  & VP-en                & $1{,}783$ \\
  & CV-en                & $1{,}000$ \\
\midrule
\multirow{6}{*}{EU}
  & CV $\times 7$ langs  & $1{,}000$ each \\
  & MLS-de               & $3{,}394$ \\
  & MLS-nl               & $3{,}075$ \\
  & MLS-fr $/$ es        & $2{,}426 / 2{,}385$ \\
  & MLS-it $/$ pt $/$ pl & $1{,}262 / 871 / 520$ \\
  & VP $\times 6$ langs  & $1{,}117$--$1{,}957$ \\
\midrule
\multirow{6}{*}{CJK}
  & CV-zh                & $1{,}000$ \\
  & AISHELL-zh           & $7{,}176$ \\
  & CV-ja                & $1{,}000$ \\
  & Reazon-ja            & $1{,}000$ \\
  & FLEURS-ja            & $650$ \\
  & Kspon-ko             & $1{,}000$ \\
\bottomrule
\end{tabular}
\caption{\textbf{Evaluation splits.}
CV splits, Reazon-ja, and Kspon-ko use a $1{,}000$-utterance
subsample (seed $42$); remaining splits use the full official
$30$s-filtered test set. The EU group covers CommonVoice
($7$ languages: de, nl, fr, es, it, pt, pl), MLS (same $7$),
and VoxPopuli ($6$; no Portuguese).}
\label{tab:splits}
\end{table}

For ReazonSpeech, which has no official train/test split, we hold
out $1{,}000$ audio paths (seed $42$, $\leq 30$\,s) as the
Reazon-ja evaluation split and blacklist the same paths from
training via \texttt{--audio\_path\_blacklist\_path}
(Appendix~\ref{app:data}). This guarantees train $\cap$ test
$= \emptyset$ on ReazonSpeech.

\subsection{Text Normalization}
\label{app:normalize}

We apply a single language-aware text normalizer at both training
(per-utterance label normalization before tokenization) and
evaluation (applied to both reference and hypothesis before
WER/CER scoring), so the train and eval distributions are
identical by construction. For English we use the HuggingFace
\texttt{EnglishTextNormalizer} (the Whisper convention), which
expands contractions, lowercases, removes ASCII punctuation, and
strips common fillers. The remaining ten languages use the
HuggingFace \texttt{BasicTextNormalizer}, which lowercases and
strips ASCII punctuation while preserving CJK characters, Hangul,
and Latin diacritics. Both routes finally collapse internal
whitespace. Using one function at both sites avoids the
train--eval normalization gap that arises when an alternative
English normalizer is used at evaluation time.

\subsection{Language-Aware MBR Metric}
\label{app:mbr}

For both the MBR pairwise distance and the oracle upper-bound
selection (\S\ref{sec:method-pdd}), we use a language-aware edit
rate $d(\cdot, \cdot)$:
\begin{equation*}
d \;=\;
\begin{cases}
\text{CER} & \text{if } \ell \in \{\text{zh, ja, ko, th, lo, my, km}\}\\
\text{WER} & \text{otherwise.}
\end{cases}
\end{equation*}
The same callable is reused for every ordered candidate pair when
computing the MBR risk and for every $(\text{ref}, \text{cand})$
pair when computing the oracle pick. Korean is included in the
CER set even though some leaderboards report Korean WER, because
Hangul word boundaries are sparse and orthographically ambiguous;
using CER for both the MBR pairwise distance and the oracle pick
recovers $1$--$3$\,pp on Korean oracle WER (CV-ko, Kspon-ko) and
$0.1$--$0.3$\,pp on Korean MBR selection in our validation runs.
For the CJK group in Table~\ref{tab:cjk-main} we therefore report
CER for every cell, including Kspon-ko, matching the convention
of recent multilingual diffusion ASR systems
\citep{navon2025drax, yen2026mdmasr}.

\begin{table*}[!t]
\centering
\small
\setlength{\tabcolsep}{4pt}
\begin{tabular}{llll}
\toprule
\textbf{Model} & \textbf{Checkpoint} & \textbf{Decoding configuration} & \textbf{Prec.} \\
\midrule
Whisper-large-v3 & \texttt{openai/whisper-large-v3}
   & greedy, $\text{beam}{=}1$, forced lang $+$ transcribe ids per utt
   & bf16 \\
Whisper-turbo & \texttt{openai/whisper-large-v3-turbo}
   & greedy, $\text{beam}{=}1$, forced lang $+$ transcribe ids per utt
   & bf16 \\
Canary-1b-v2 & \texttt{nvidia/canary-1b-v2}
   & greedy, $\text{beam}{=}1$, per-utt \texttt{target\_lang}, PnC on
   & bf16 \\
Qwen3-ASR-1.7B & \texttt{Qwen/Qwen3-ASR-1.7B}
   & \texttt{model.transcribe()}, batch $1$, full-name lang per utt
   & bf16 \\
\midrule
OWSM-CTC v3.1 & \texttt{espnet/owsm\_ctc\_v3.1\_1B}
   & CTC greedy, per-utt language symbol
   & bf16 \\
MMS-all & \texttt{facebook/mms-1b-all}
   & CTC greedy, per-utt language-adapter swap
   & bf16 \\
\midrule
Drax (MBR, $8/16$) & \texttt{aiola/drax-v1}
   & DFM, $8$ sampling steps, $K{=}16$, $\tau{=}0.10$, MBR selection
   & bf16 \\
Whisper-LLaDA
   & \makecell[tl]{\texttt{whisper-large-v3} enc\\
                   $+$ \texttt{LLaDA-8B-Instruct} $+$ paper ckpt}
   & standalone (\texttt{decoding}) mode, $128$ steps, gen len $=128$
   & fp16$/$fp32 \\
\midrule
Whisfusion (ours) & --- (this work)
   & PDD, $K{=}5$, $N{=}3$, $\tau{=}0.1$, MBR selection
   & bf16 \\
\bottomrule
\end{tabular}
\caption{\textbf{Baseline configurations.} All systems take the
same $16$\,kHz mono audio clipped to $30$\,s but use their native
preprocessing frontend; evaluation splits, text normalization,
and RTFx timing protocol are identical.}
\label{tab:baselines}
\end{table*}

\subsection{RTFx Measurement Protocol}
\label{app:rtfx}

\paragraph{Hardware.}
All runtime numbers are measured on a single NVIDIA H100 $80$\,GB
GPU (single shard, no multi-GPU contention) in bf16 precision for
both encoder and decoder, without \texttt{torch.compile} or CUDA
Graphs. The only exception is Whisper-LLaDA, which is run in
fp16$/$fp32 mixed precision following its upstream model card.

\paragraph{Audio benchmark set.}
RTFx is reported on a length-balanced $336$-utterance English
benchmark built by uniformly sampling $56$ utterances within each
of six $5$-second duration bins ($0$--$5$, $5$--$10$, \ldots,
$25$--$30$\,s), drawn from LibriSpeech test-clean and test-other,
Earnings-22, and VoxPopuli-en (mean duration $14.74$\,s). This
length-balanced design follows the runtime protocol of Drax
\citep{navon2025drax} and prevents length-skewed test sets from
favoring models whose latency does not scale with output length.

\paragraph{Timing methodology.}
Per-utterance wall time is captured with
\texttt{torch.cuda.synchronize()} on both boundaries and
\texttt{time.perf\_counter()} for the elapsed measurement. The
synchronize boundaries avoid the systematic underestimation that
would arise from asynchronous CUDA kernel launches. Corpus RTFx
follows the formula in \S\ref{sec:setup-metrics}, identical to
the Open ASR Leaderboard
\citep{srivastav2023openasrleaderboard}, Drax, and MDM-ASR.

\paragraph{Timed regions and exclusions.}
The timer covers the encoder forward pass, the decoder/PDD
forward passes, candidate generation, MBR pairwise selection,
and Whisper reranking when used. File I/O (audio load from
disk), one-time model loading, and the first $5$ warmup
utterances are excluded from the aggregate. For systems whose
public API does not expose a pre-mel entry point (Canary,
Qwen3-ASR), mel preprocessing is necessarily inside the timer;
this contributes roughly $5$--$10$\,ms per utterance, small
relative to the RTFx ranges we report.

\paragraph{Baseline re-measurement.}
All baselines reported in Tables~\ref{tab:en-main},
\ref{tab:eu-main}, and \ref{tab:cjk-main} are re-measured on the
same H100 under this protocol; we do not reuse RTFx values
published in baseline papers, since absolute RTFx is hardware-
and precision-dependent.

\subsection{Baseline Configurations}
\label{app:baselines}

Table~\ref{tab:baselines} lists the exact checkpoint and decoding
configuration used for each baseline in the main results. All
baselines take the same $16$\,kHz mono audio input clipped to at
most $30$\,s; each model then uses its native frontend and
official preprocessing path (HF \texttt{WhisperFeatureExtractor}
for Whisper-large-v3, Whisper-turbo, Drax, and Whisper-LLaDA;
the NeMo FastConformer preprocessor for Canary-1b-v2; the ESPnet
preprocessor for OWSM-CTC; the wav2vec2 frontend for MMS-all;
and the audio encoder built into Qwen3-ASR). Evaluation splits,
the language-aware text normalizer of Appendix~\ref{app:normalize},
and the RTFx timing protocol of Appendix~\ref{app:rtfx} with batch
size $1$ on a single H100 are identical across systems.

Three points to note. \textbf{Drax}: we report the
\texttt{mbr\_8\_16} operating point ($8$ sampling steps, $K{=}16$,
MBR pairwise selection), matching the headline column of the Drax
paper \citep{navon2025drax}; the single-shot and Whisper-rescore
Drax variants are not reported. \textbf{Whisper-LLaDA}: we use the
standalone (non-cascade) decoding mode, which takes no external
transcript as input. The paper's cascade mode refines an
externally produced Whisper-LLaMA transcript and is excluded for a
single-model NAR comparison. \textbf{Canary}: we use the
\texttt{1b-v2} checkpoint; the earlier \texttt{1b-flash} checkpoint
covers only four languages (en/de/es/fr) and is incompatible with
our $11$-language evaluation. Precision is bf16 for all baselines
except Whisper-LLaDA, where LLaDA-8B is loaded in fp16 and the
Whisper encoder runs in fp32, following the upstream model card.

\section{Hyperparameter Sensitivity}
\label{app:hparams}

All sweeps in this section use the same five English splits as
Figure~\ref{fig:k-sweep} (LS test-clean$/$-other, CV-en,
Earnings-22, VoxPopuli-en) and report MBR WER at $K{=}5$ unless
otherwise stated. Avg is the macro-average over the five splits.

\subsection{Denoising Steps $N$}
\label{app:hparam-N}

Table~\ref{tab:n-sweep} sweeps the number of denoising steps $N$
with $K{=}5$ and $\tau{=}0.1$ fixed; the mask-ratio trajectory is
extended in the high-mask pattern to match each $N$. Increasing
$N$ from $1$ to $2$ recovers $5.3$\,pp on average; the
$N{=}2 \to 3$ gain is $0.6$\,pp; and $N{=}3 \to 5$ yields only an
additional $0.3$\,pp at twice the decoder cost. We therefore
adopt $N{=}3$ as the default operating point.

\begin{table}[h]
\centering
\small
\setlength{\tabcolsep}{4pt}
\begin{tabular}{ccccccc}
\toprule
$N$ & \textbf{Avg} & LS-c & LS-o & CV-en & E22 & VP-en \\
\midrule
$1$               & $12.46$ & $3.92$ & $8.61$ & $15.38$ & $18.75$ & $15.64$ \\
$2$               &  $7.18$ & $1.97$ & $4.21$ & $10.82$ & $11.38$ &  $7.50$ \\
$\mathbf{3}$ (def.) & $\mathbf{6.55}$ & $1.74$ & $3.78$ & $10.25$ & $10.33$ &  $6.66$ \\
$4$               &  $6.29$ & $1.71$ & $3.64$ &  $9.60$ &  $9.99$ &  $6.51$ \\
$5$               &  $6.22$ & $1.70$ & $3.67$ &  $9.51$ &  $9.83$ &  $6.39$ \\
\bottomrule
\end{tabular}
\caption{\textbf{Denoising-step sweep ($N$).}
MBR WER on the five English splits with $K{=}5$, $\tau{=}0.1$.
\textbf{LS-c}$/$\textbf{LS-o}: LibriSpeech test-clean$/$-other;
\textbf{E22}: Earnings-22.}
\label{tab:n-sweep}
\end{table}

\subsection{Sampling Temperature $\tau$}
\label{app:hparam-tau}

Table~\ref{tab:tau-sweep} sweeps the softmax sampling temperature
$\tau$ with $K{=}5$, $N{=}3$. MBR WER is flat across
$\tau \in [0.05, 0.30]$, all within $\pm 0.02$\,pp; this band is at
or below the run-to-run RTFx noise floor of the bench
(Appendix~\ref{app:rtfx}). We use $\tau{=}0.1$ as the default
because it lies in the center of this flat region.

\begin{table}[h]
\centering
\small
\setlength{\tabcolsep}{4pt}
\begin{tabular}{ccccccc}
\toprule
$\tau$ & \textbf{Avg} & LS-c & LS-o & CV-en & E22 & VP-en \\
\midrule
$0.05$               & $6.54$ & $1.76$ & $3.80$ & $10.16$ & $10.35$ & $6.63$ \\
$\mathbf{0.10}$ (def.) & $\mathbf{6.55}$ & $1.74$ & $3.78$ & $10.25$ & $10.33$ & $6.66$ \\
$0.20$               & $6.53$ & $1.73$ & $3.78$ & $10.16$ & $10.31$ & $6.66$ \\
$0.30$               & $6.56$ & $1.73$ & $3.80$ & $10.18$ & $10.31$ & $6.79$ \\
\bottomrule
\end{tabular}
\caption{\textbf{Temperature sweep ($\tau$).}
MBR WER on the five English splits with $K{=}5$, $N{=}3$.}
\label{tab:tau-sweep}
\end{table}

\subsection{Trajectory Schedule}
\label{app:hparam-traj}

Table~\ref{tab:sched-sweep} compares five mask-ratio trajectories
spanning the design space, where the last-step input mask ratio
controls how much context is exposed for the final commit. The
sweet spot is a high-mask trajectory with a $60$--$80\%$ final
mask ratio. The two extremes degrade significantly: a too-narrow
trajectory ($90\%$ final) starves intermediate steps of textual
context ($+0.55$\,pp); a too-aggressive linear or low-final
trajectory ($33\%$ and $10\%$) commits too little mass at the
final step ($+0.66$ and $+2.92$\,pp, respectively). This directly
supports the high-mask design choice of \S\ref{sec:method-pdd}.

\begin{table}[h]
\centering
\small
\setlength{\tabcolsep}{4pt}
\begin{tabular}{lcc}
\toprule
\textbf{Trajectory} & \textbf{Last mask} & \textbf{Avg} \\
\midrule
$[1.0, 0.95, 0.9]$  & $90\%$ & $7.10$ \\
$[\mathbf{1.0, 0.9, 0.8}]$ (def.) & $\mathbf{80\%}$ & $\mathbf{6.55}$ \\
$[1.0, 0.8, 0.6]$   & $60\%$ & $6.52$ \\
$[1.0, 0.667, 0.333]$ & $33\%$ & $7.21$ \\
$[1.0, 0.4, 0.1]$   & $10\%$ & $9.47$ \\
\bottomrule
\end{tabular}
\caption{\textbf{Trajectory schedule sweep.}
MBR WER on the five English splits with $K{=}5$, $N{=}3$,
$\tau{=}0.1$. \textbf{Last mask} is the input mask ratio at the
final denoising step.}
\label{tab:sched-sweep}
\end{table}

\section{Per-split European Results}
\label{app:per-split}

Table~\ref{tab:eu-per-split} reports per-split WER for all $20$
European benchmarks, expanding the family averages of
Table~\ref{tab:eu-main}. Drax does not provide outputs for Dutch
or Polish ($6$ of $20$ splits); its average is computed over the
remaining $14$ splits.

\begin{table*}[!ht]
\centering
\scriptsize
\setlength{\tabcolsep}{2.5pt}
\resizebox{\textwidth}{!}{%
\begin{tabular}{lcccccccccccccccccccccc}
\toprule
\multirow{2}{*}{\textbf{Model}}
 & \multicolumn{3}{c}{\textbf{DE}}
 & \multicolumn{3}{c}{\textbf{NL}}
 & \multicolumn{3}{c}{\textbf{FR}}
 & \multicolumn{3}{c}{\textbf{ES}}
 & \multicolumn{3}{c}{\textbf{IT}}
 & \multicolumn{2}{c}{\textbf{PT}}
 & \multicolumn{3}{c}{\textbf{PL}}
 & \multirow{2}{*}{\textbf{Avg}} \\
\cmidrule(lr){2-4}\cmidrule(lr){5-7}\cmidrule(lr){8-10}\cmidrule(lr){11-13}\cmidrule(lr){14-16}\cmidrule(lr){17-18}\cmidrule(lr){19-21}
 & MLS & CV & VP & MLS & CV & VP & MLS & CV & VP
 & MLS & CV & VP & MLS & CV & VP & MLS & CV
 & MLS & CV & VP & \\
\midrule
Whisper-large-v3   & 5.6 & 5.1 & 17.1 & \underline{10.3} & 5.1 & 22.6 & 4.8 & 11.2 & 10.1 & \underline{4.1} & 4.7 & 10.2 & \textbf{9.3} & 5.9 & 27.8 & 7.2 & 6.1 & \textbf{4.4} & 5.7 & \underline{8.3} & 9.3 \\
Whisper-turbo      & 6.3 & 7.1 & 21.1 & 10.4 & 6.7 & 27.9 & 5.1 & 12.1 & 11.3 & 4.3 & 5.4 & 15.4 & 9.9 & 6.4 & 31.2 & \textbf{5.7} & 6.9 & \underline{4.6} & 6.6 & 9.8 & 10.7 \\
Canary-1b-v2       & \textbf{4.9} & 5.6 & \textbf{9.5} & 10.4 & 4.9 & \underline{11.5} & \textbf{4.2} & \textbf{7.8} & \textbf{8.8} & \textbf{3.4} & \underline{4.3} & \textbf{7.4} & 11.2 & \textbf{4.8} & \textbf{14.4} & 8.0 & 11.1 & 7.6 & 5.6 & \textbf{7.4} & \textbf{7.6} \\
Qwen3-ASR-1.7B     & 5.9 & 5.5 & 11.5 & 12.2 & 5.8 & 14.1 & 5.3 & \textbf{7.8} & 9.3 & 4.6 & 4.4 & 7.7 & 13.3 & 5.0 & 17.0 & 7.7 & 6.9 & 15.3 & 14.7 & 13.9 & 9.4 \\
OWSM-CTC v3.1      & 11.9 & 11.6 & 16.4 & 20.3 & 19.4 & 26.6 & 13.0 & 16.0 & 15.7 & 10.4 & 11.9 & 13.6 & 22.6 & 16.4 & 25.4 & 23.9 & 21.8 & 31.8 & 31.7 & 29.6 & 19.5 \\
MMS-all            & 8.8 & 11.8 & 13.0 & 12.8 & 10.1 & 14.5 & 8.8 & 15.9 & 12.1 & 5.8 & 9.4 & 9.6 & 11.0 & 9.9 & 17.2 & 16.3 & 14.0 & 10.6 & 9.3 & 9.4 & 11.5 \\
Drax (MBR, 8/16)   & 7.0 & 7.9 & 11.0 & --- & --- & --- & 6.5 & 11.2 & 10.5 & 4.8 & 6.2 & 8.7 & 11.0 & 8.1 & 16.2 & 11.6 & 12.1 & --- & --- & --- & 9.5$^{\dagger}$ \\
\midrule
Whisfusion (ours)  & 6.6 & 5.4 & 10.4 & 11.3 & \underline{3.2} & 13.1 & 5.2 & 8.6 & 9.7 & 5.0 & 4.9 & 8.4 & 10.6 & 5.3 & 16.8 & 13.2 & \underline{3.8} & 9.8 & \underline{3.3} & 9.3 & 8.2 \\
\,\,$+$ Whisper rerank & \underline{6.2} & \textbf{4.9} & \underline{10.0} & \textbf{10.9} & \textbf{2.8} & \textbf{12.7} & \underline{4.9} & \textbf{7.8} & \underline{9.6} & 4.6 & \textbf{4.2} & \underline{8.1} & \underline{9.8} & \underline{4.8} & \underline{16.3} & \underline{12.4} & \textbf{3.4} & 8.7 & \textbf{3.0} & 8.9 & \underline{7.7} \\
\bottomrule
\end{tabular}%
}
\caption{\textbf{Per-split European WER (\%).}
All numbers rounded to one decimal place.
$^{\dagger}$Drax average is computed over the $14$ of $20$
splits where it provides outputs (Dutch and Polish are
unsupported).}
\label{tab:eu-per-split}
\end{table*}

\section{Decoding Strategy Analysis}
\label{app:decoding}

\subsection{Random vs.\ Confidence-based Remasking}
\label{app:remask}

This appendix backs up the two qualitative claims made in
\S\ref{sec:method-pdd} for choosing random Bernoulli remasking
over confidence-based selection: (i) confidence-based remasking
degrades final WER, and (ii) high-mask-regime confidences are
poorly calibrated.

\paragraph{(i) WER comparison.}
Table~\ref{tab:remask} replaces the random Bernoulli remasking
step in PDD with a confidence-based variant that keeps the
top-$\lfloor (1 - r_{n+1}) L \rfloor$ tokens by softmax confidence
and re-masks the rest, holding all other settings fixed
($K{=}5$, $N{=}3$, $\tau{=}0.1$, default trajectory). Random
remasking is $1.26$\,pp better in MBR WER on average. The
confidence variant also yields essentially identical $\text{sel}$
and MBR WER ($7.78$ vs.\ $7.81$), indicating that the $K$
candidates become near-duplicates and MBR consensus provides no
benefit, in contrast to the $0.30$\,pp consensus gain we see
under random remasking ($6.85 \to 6.55$).

\begin{table}[h]
\centering
\small
\setlength{\tabcolsep}{3pt}
\resizebox{\columnwidth}{!}{%
\begin{tabular}{lcccccc}
\toprule
\textbf{Remask} & \textbf{Avg} & LS-c & LS-o & CV-en & E22 & VP-en \\
\midrule
\textbf{Random} (def.) & $\mathbf{6.55}$ & $1.74$ & $3.78$ & $10.25$ & $10.33$ & $6.66$ \\
Confidence            & $7.81$ & $2.27$ & $4.79$ & $11.49$ & $11.86$ & $8.64$ \\
$\Delta$              & $+1.26$ & $+0.53$ & $+1.02$ & $+1.24$ & $+1.53$ & $+1.97$ \\
\bottomrule
\end{tabular}%
}
\caption{\textbf{Random vs.\ confidence-based remasking.}
MBR WER on the five English splits with $K{=}5$, $N{=}3$,
$\tau{=}0.1$.}
\label{tab:remask}
\end{table}

\begin{table}[!t]
\centering
\small
\setlength{\tabcolsep}{3pt}
\resizebox{\columnwidth}{!}{%
\begin{tabular}{ccccc}
\toprule
$t$ & Accuracy & Conf.\ & ECE & Conf.\ $-$ Acc.\ \\
\midrule
$0.5$ & $99.30$\% & $99.17$\% & $0.0026$ & $-0.13$ (under) \\
$0.7$ & $99.05$\% & $98.91$\% & $0.0034$ & $-0.14$ (under) \\
$0.9$ & $97.89$\% & $98.22$\% & $0.0035$ & $+0.32$ (over) \\
$\mathbf{1.0}$ & $\mathbf{92.38}$\% & $\mathbf{93.52}$\% & $\mathbf{0.0239}$ & $\mathbf{+1.14}$ (over) \\
\bottomrule
\end{tabular}%
}
\caption{\textbf{Decoder confidence calibration at masked positions
versus mask ratio $t$.} ECE is the expected calibration error
($15$ confidence bins). At $t{=}1.0$ ECE jumps roughly $9\times$
relative to $t{=}0.5$.}
\label{tab:calib}
\end{table}

\paragraph{(ii) Calibration at high mask ratios.}
Table~\ref{tab:calib} probes the decoder's confidence calibration
at the masked positions. We corrupt LibriSpeech test-clean
references at a fixed mask ratio $t$, run a single decoder forward
($200$ utterances $\times$ $4$ Monte-Carlo mask realizations),
and bin the predictions at masked positions by softmax
confidence ($15$ bins). At moderate mask ratios ($t{=}0.5$,
$0.7$) the decoder is essentially perfectly calibrated, with
expected calibration error (ECE) below $0.004$. As $t$ approaches
$1$, the model becomes increasingly over-confident: at the fully
masked regime ($t{=}1.0$) ECE jumps to $0.0239$
(${\sim}9\times$), with confidence over-predicting accuracy by
$1.14$\,pp. The implication is direct: confidence-based remasking
preferentially commits high-confidence positions, which at
$t{\approx}1$ are also the most over-confident positions, so
errors are committed early and propagate. Random Bernoulli
remasking sidesteps this miscalibration entirely.

\end{document}